\documentclass{jfm}

\usepackage{graphicx}
\usepackage{natbib}
\usepackage{amsmath,amstext,amssymb}
\usepackage{subfigure}

\ifCUPmtlplainloaded \else
  \checkfont{eurm10}
  \iffontfound
    \IfFileExists{upmath.sty}
      {\typeout{^^JFound AMS Euler Roman fonts on the system,
                   using the 'upmath' package.^^J}%
       \usepackage{upmath}}
      {\typeout{^^JFound AMS Euler Roman fonts on the system, but you
                   dont seem to have the}%
       \typeout{'upmath' package installed. JFM.cls can take advantage
                 of these fonts,^^Jif you use 'upmath' package.^^J}%
      }
  \else
  \fi
\fi

\ifCUPmtlplainloaded \else
  \checkfont{msam10}
  \iffontfound
    \IfFileExists{amssymb.sty}
      {\typeout{^^JFound AMS Symbol fonts on the system, using the
                'amssymb' package.^^J}%
       \usepackage{amssymb}%
         
         \let\geq=\geqslant
      }{}
  \fi
\fi

\ifCUPmtlplainloaded \else
  \IfFileExists{amsbsy.sty}
    {\typeout{^^JFound the 'amsbsy' package on the system, using it.^^J}%
     \usepackage{amsbsy}}
    {\providecommand\boldsymbol[1]{\mbox{\boldmath $##1$}}}
\fi

\newcommand\bU{\mathbf{U}}
\newcommand\bu{\mathbf{u}}
\newcommand\curl{\nabla\times}

\title[Faraday instability on a sphere: Floquet analysis]{Faraday instability on a sphere: \\Floquet analysis}

\author[A. Ebo-Adou, L.S.~Tuckerman]
       {A. Ebo-Adou$^{1,2}$, L.S.~Tuckerman$^1$
\thanks{Email address for correspondence: laurette@pmmh.espci.fr}}

\affiliation{
  $^1$Laboratoire de Physique et M\'ecanique des Milieux H\'et\'erog\`enes (PMMH), UMR CNRS 7636 ; PSL-ESPCI;
  Sorbonne Univ.- UPMC, Univ.~Paris 6; Sorbonne Paris Cit\'e-UDD, Univ.~Paris 7
  \\[\affilskip]
$^2$LIMSI, CNRS, Universit\'e Paris-Saclay, 91405 Orsay, France}

\pubyear{}
\volume{}
\pagerange{}
\begin{document}

\maketitle

%Abstract
%%%%%%%%%%%%%%%%%%%%%%%%%%%%%%%%%%%%%%%%%%%%%%%%%%%%%%%%%%%%%%%%%%%%%%%%%%%%%%%%%%%%%%%%%%%%%%%%%%%%%%%%%%%%%%%%%%%%%%%%%%%%%%%%%%%%%%%%%%%%%%%%%%%%%%%%%%%%%%%%%%%%%%
\begin{abstract}

Standing waves appear at the surface of a spherical viscous liquid
drop subjected to radial parametric oscillation. This is the spherical
analogue of the Faraday instability.  Modifying the \cite{KT1994}
planar solution to a spherical interface, we linearize the governing
equations about the state of rest and solve the resulting equations by
using a spherical harmonic decomposition for the angular dependence,
spherical Bessel functions for the radial dependence, and a Floquet
form for the temporal dependence. Although the inviscid
problem can, like the planar case, be mapped exactly onto the Mathieu
equation, the spherical geometry introduces additional terms into
the analysis.
The dependence of the threshold on viscosity is studied and
scaling laws are found. 
It is shown that the spherical thresholds are similar to
the planar infinite-depth thresholds, even for small
wavenumbers for which the curvature is high.
A representative time-dependent Floquet mode is displayed.

\end{abstract}
{\it published in Journal of Fluid Mechanics {\bf 805}, 591-610 (2016)}

%%%%%%%%%%%%%%%%%%%%%%%%%%%%%%%%%%%%%%%%%%%%%%%%%%%%%%%%%%%%%%%%%%%%%%%%%%%%%%%%%%%%%%%%%%%%%%%%%%%%%%%%%%%%%%%%%%%%%%%%%%%%%%%%%%%%%%%%%%%%%%%%%%%%%%%%%%%%%%%%%%%%%%
\section{Introduction}
%%%%%%%%%%%%%%%%%%%%%%%%%%%%%%%%%%%%%%%%%%%%%%%%%%%%%%%%%%%%%%%%%%%%%%%%%%%%%%%%%%%%%%%%%%%%%%%%%%%%%%%%%%%%%%%%%%%%%%%%%%%%%%%%%%%%%%%%%%%%%%%%%%%%%%%%%%%%%%%%%%%%%%

The dynamics of oscillating drops are of interest to researchers in
pattern formation and dynamical systems as well as having practical
applications over a wide variety of scales, in areas as diverse as
astroseismology, containerless material processing for high purity
crystal growth and drug delivery and mixing in microfluidic devices.

Surface tension is responsible for the spherical shape of a drop.  In
the absence of external forces, if the drop is slightly perturbed, it
will recover its spherical shape through decaying oscillations.  This
problem was first considered by \cite{Kel1863} and \cite{Rayl1879},
who described natural oscillations of drops of inviscid fluids.
\cite{Rayl1879} and \cite{Lamb1932} derived the now-classic resonance
mode frequency resulting from the restoring force of surface tension:
\begin{equation}
\omega^2 = \frac{\sigma}{\rho} \frac{\ell(\ell-1)(\ell+2)}{R^3}
\label{eq:Lamb}  \end{equation}
where $\omega$ is the frequency, $\sigma$ and $\rho$ the surface
tension and density, $R$ is the radius, and $\ell$ is
the degree of the spherical harmonic
\begin{equation}
Y_\ell^m=P_\ell^m(\cos\theta)e^{im\phi}
\label{eq:ylm}  \end{equation}
describing the perturbation.
Linear analyses including viscosity were carried out by \cite{Reid1960},
\cite{Chan1961} and \cite{MS1968}.  These authors
demonstrated the equivalence of this problem to that of a fluid globe
oscillating under the influence of self-gravitation, generalizing the
previous conclusion of Lamb.  Chandrasekhar showed that the return to
a spherical shape could take place via monotonic decay as well as via
damped oscillations. The problem was further investigated 
by \cite{Pros1980} using an initial-value code.
Weakly nonlinear effects in inviscid fluid drops were investigated by
\cite{TB1983} using a Poincar\'e-Lindstedt expansion technique.

The laboratory realization of any configuration with only spherically
symmetric radially directed forces is difficult.  Indeed such
experiments have been sent into space, e.g. 
\citep{WAL1996,futterer2013sheet} and in parabolic flight
\citep{Fal2009} in order to eliminate or
reduce the perturbing influence of the gravitational field of the
Earth.  \cite{WAL1996} were able to confirm the decrease in frequency
with increasing oscillation amplitude predicted by \cite{TB1983}.
\cite{WAL1996} mention, however, that the treatment of viscosity is not 
exact. \cite{Fal2009} produced spherical capillary wave
turbulence and compared its spectrum with theoretical predictions.

In the laboratory, drops have been levitated by using acoustic or
magnetic forces and excited by periodic electric modulation
\citep{SXW2010}; drops and puddles weakly pinned on a vibrating
substrate \citep{BS2011} have produced star-like patterns.  One of the
purposes of such experiments is to provide a measurement of the
surface tension.  \cite{TZW1982} visualized the shapes and internal
flow of vibrating drops and compared the frequencies to those of
\cite{Lamb1932} and the damping coefficients to those derived by
\cite{Mars1980}.  These experimental procedures cannot produce a
perfectly spherical base state, and indeed, \cite{TW1982} and
\cite{CB1991} discuss differences between oscillating oblate and
prolate drops, and the resulting deviations from \eqref{eq:Lamb}.
Because the experimentally observed frequencies remain close
to \eqref{eq:Lamb}, it seems likely that the results
of our stability analysis are also only mildly affected by a departure
from perfect spherical symmetry.

Here, and in a companion paper, we consider a viscous drop under the
influence of a time-periodic radial bulk force and of surface tension.
Our investigation relies on a variety of mathematical and
computational tools.  Here, we solve the linear stability problem by
adapting to spherical coordinates the Floquet method of \cite{KT1994}.
At the linear level, the instability depends only on the spherical
wavenumber $\ell$ of \eqref{eq:ylm} as illustrated by the Lamb-Rayleigh
relation \eqref{eq:Lamb}.
%Thus, perturbations can be assumed to be
%axisymmetric without loss of generality.
Thus, perturbations which are not axisymmetric ($m\neq 0$ in
\eqref{eq:ylm}) have the same thresholds as the corresponding
axisymmetric ($m=0$) perturbations.  Indeed, the theoretical and
numerical investigations listed above have assumed that the drop shape
remains axisymmetric.  The fully nonlinear Faraday problem, however,
usually leads to patterns which are non-axisymmetric.  In our
complementary investigation \citep{Ali2}, we will describe the results
of full three-dimensional simulations which calculate the interface
motion and the velocity field inside and outside the parametrically
forced drop and interpret them in the context of the theory of pattern
formation.

%%%%%%%%%%%%%%%%%%%%%%%%%%%%%%%%%%%%%%%%%%%%%%%%%%%%%%%%%%%%%%%%%%%%%%%%%%%%%%%%%%%%%%%%%%%%%%%%%%%%%%%%%%%%%%%%%%%%%%%%%%%%%%%%%%%%%%%%%%%%%%%%%%%%%%%%%%%%%%%%%%%%%%
\section{Governing equations}
%%%%%%%%%%%%%%%%%%%%%%%%%%%%%%%%%%%%%%%%%%%%%%%%%%%%%%%%%%%%%%%%%%%%%%%%%%%%%%%%%%%%%%%%%%%%%%%%%%%%%%%%%%%%%%%%%%%%%%%%%%%%%%%%%%%%%%%%%%%%%%%%%%%%%%%%%%%%%%%%%%%%%%

%%%%%%%%%%%%%%%%%%%%%%%%%%%%%%%%%%%%%%%%%%%%%%%%%%%%%%%%%%%%%%%%%%%%%%%%%%%%%%%%%%%%%%%%%%%%%%%%%%%%%%%%%%%%
\subsection{Equations of motion}
%%%%%%%%%%%%%%%%%%%%%%%%%%%%%%%%%%%%%%%%%%%%%%%%%%%%%%%%%%%%%%%%%%%%%%%%%%%%%%%%%%%%%%%%%%%%%%%%%%%%%%%%%%%%
We consider a drop of viscous, incompressible liquid bounded by a spherical free surface that separates the liquid from the 
exterior in the presence of an uniform radial oscillatory body force. The fluid motion inside the drop satisfies the Navier-Stokes equations

\begin{subequations}\begin{align}
    \rho \left[ \frac{\partial}{\partial t} + ( \bU \cdot \nabla )  \right] \bU  & = - \nabla P + \eta \nabla^2 \bU - \rho G(r,t) \vec{e}_r     \label{NS}\\
        \nabla \cdot \bU &= 0 \nonumber
\end{align}\end{subequations}
where $\bU$ is the velocity, $P$ the pressure, 
$\rho$ the density and $\eta$ the dynamic viscosity. 
$G(r,t)$ is an imposed radial parametric acceleration given by 
\begin{equation}
G(r,t) = \left( g - a \cos(\omega t) \right) \frac{r}{R}  % \mbox{ with $b$ an integer.}
\label{eq:Gdef}\end{equation}
which is regular at the origin.
%and whose form has been chosen for its simplicity.

\noindent 
The interface is located at 
\begin{equation}
r=R+\zeta(\theta,\phi,t)
\end{equation}
Conservation of volume leads to the requirement that the integral of 
$\zeta$ over the sphere must be zero. 
%
%%%%%%%%%%%%%%%%%%%%%%%%%%%%%%%%
%\subsection{Boundary condition}%
%%%%%%%%%%%%%%%%%%%%%%%%%%%%%%%%

\noindent The boundary conditions applied at the interface are the kinematic condition, which states that the interface is advected by the fluid

\begin{equation}
\left[ \frac{\partial}{\partial t}  + \left( \bU \cdot \nabla \right) \right] 
\zeta   = U_r |_{r=R+\zeta}
\label{kincondg}
\end{equation}

%\subsubsection{The Stress Balance Equation}
%%%%%%%%%%%%%%%%%%%%%%%%%%%%%%%%%%%%%%%%%
\noindent and the interface stress balance equation, which is given by 

\begin{equation}
\mathbf{n} \cdot \mathbf{\hat{\Pi}} - \mathbf{n} \cdot \mathbf{\Pi}  = \sigma \mathbf{n}(\nabla \cdot \mathbf{n}) - \nabla \sigma
\label{sbr}
\end{equation}

\noindent 
where $\sigma$ is the surface tension coefficient and $\mathbf{n}$ represents the unit outward normal to the surface, both defined only on the interface.
\noindent The tensors $\mathbf{\Pi}$ (drop) and $\mathbf{\hat{\Pi}}$
(medium) denote the stress tensor in each fluid and are defined by

\begin{equation}
\mathbf{\Pi} = -P \mathbf{I_d} + \eta \left[ \nabla \bU + \left( \nabla \bU \right)^T \right].
\label{Pidef}\end{equation}

\noindent

For simplicity, we consider a situation in which the outer medium has
no effect on the drop, and so we set the density, pressure and stress
tensor $\mathbf{\hat{\Pi}}$ outside the drop to zero. The boundary
conditions corresponding to the case of drop forced in a medium are
described in the appendix. We assume that the surface tension is
uniform, so $\nabla \sigma = 0$. The tangential stress balance
equation at the free surface then reduces to

\begin{equation}
\mathbf{n} \cdot \mathbf{\Pi}\cdot\mathbf{t} = 0
 \label{tstb}
\end{equation}
%\noindent
for both unit tangent vectors $\mathbf{t}$. 

The normal stress jump boundary condition determines the curvature of the deformed interface.
The Laplace formula relates the normal stress jump to the divergence
of the normal field, which is in turn equal to the mean curvature:
\begin{equation}
- \mathbf{n} \cdot \mathbf{\Pi} \cdot \mathbf{n}  = 
\sigma \nabla \cdot \mathbf{n}  
= \sigma \left( \frac{1}{R_1} + \frac{1}{R_2} \right)
\label{nsb}
\end{equation}
with $R_1$ and $R_2$ the principal radii of curvature at a given point of the surface.

For a sphere, $R_1=R_2=R$ and equation \eqref{nsb} becomes 
\begin{equation}
P|_{r=R} = 2\frac{\sigma}{R}
\end{equation}
and the solution to the governing equations \eqref{NS} and boundary conditions
\eqref{kincondg}, \eqref{tstb}, \eqref{nsb} is the motionless equilibrium
spherical state at $r=R$ with 
\begin{subequations}\begin{align}
   \bar{\bU} & = 0   \\
   \bar{P}(r,t) & = 2 \frac{\sigma}{R} - \int_r^R \rho G(r',t) dr'  
\end{align}\label{BF}\end{subequations}
where $\bar{P}$ is continuously differentiable at the origin because 
$G(0,t)=0$.

%%%%%%%%%%%%%%%%%%%%%%%%%%%%%%%%%%%%%%%%%%%%%%%%%%%%%%%%%%%%%%%%%%%%%%%%%%%%%%%%%%%%%%%%%%%%%%%%%
\subsection{Linearizing the governing equations}
%%%%%%%%%%%%%%%%%%%%%%%%%%%%%%%%%%%%%%%%%%%%%%%%%%%%%%%%%%%%%%%%%%%%%%%%%%%%%%%%%%%%%%%%%%%%%%%%%%

%\subsubsection{Navier-Stokes Equation} 
%%%%%%%%%%%%%%%%%%%%%%%%%%%%%%%%%%%%%%%

We linearize the Navier-Stokes equations about the unperturbed state (\ref{BF}) by decomposing the velocity and the pressure 
\begin{subequations}\begin{align}
	\bU & = \bar{\bU} + \bu \\
	P & = \bar{P} + p
\end{align}\end{subequations}
\noindent which leads to the equation for the perturbation fields $\bu$ and $p$
\begin{subequations}    \begin{align}
        \rho\frac{ \partial \bu}{\partial t}  &= - \nabla p + \eta \nabla^2 \bu \label{fhs} \\
        \nabla \cdot \bu &= 0 \label{fhs_inc}
    \end{align}\end{subequations}
We write the definitions in spherical coordinates of various differential operators: 
\begin{subequations}
\begin{align}
\nabla_H \cdot & \equiv \frac{1}{r \sin \theta} \frac{\partial 
}{\partial \theta} \sin \theta \, \hat{e}_\theta\cdot + \frac{1}{r \sin \theta} \frac{\partial }{\partial \phi} \hat{e}_\phi\cdot\\ 
 \nabla_H  & \equiv \frac{\hat{e}_\theta}{r }\frac{\partial}{\partial\theta} + 
\frac{\hat{e}_\phi}{r \sin \theta} \frac{\partial}{\partial \phi} \\ 
\nabla^2_H & \equiv \frac{1}{r² \sin \theta} \frac{\partial}{\partial \theta} \left( \sin \theta \frac{\partial}{\partial \theta} \right) + \frac{1}{r² \sin² \theta} \frac{\partial²}{\partial \phi²} 
\label{Op_Sph}
\end{align}
For a solenoidal field satisfying \eqref{fhs_inc}, definitions \eqref{Op_Sph} lead to 
\begin{align}
  (\mathbf{\nabla}^2\bu)_r &=\left(\frac{1}{r^2}\frac{\partial}{\partial r} r^2\frac{\partial}{\partial r} + \nabla^2_H-\frac{2}{r^2}\right)u_r -\frac{2}{r}\nabla_H\cdot\bu_H \nonumber\\
 & =\left(\frac{1}{r^2}\frac{\partial}{\partial r} r^2\frac{\partial}{\partial r} -\frac{2}{r^2} +\frac{2}{r^3}\frac{\partial}{\partial r} r^2  + \nabla^2_H \right)u_r  \nonumber\\
 & =\left(\frac{1}{r^3}\frac{\partial}{\partial r} r^2\frac{\partial}{\partial r}r  + \nabla^2_H \right)u_r 
\equiv\widetilde{\mathcal{L}}^2 u_r
\end{align}
\end{subequations}
\noindent We can then eliminate the horizontal velocity ${\bf u}_H=(u_\theta ,u_\phi )$ and the pressure $p$ from \eqref{fhs} in the usual way by operating with $\vec{e}_r \cdot \curl \curl$ on \eqref{fhs}, leading to:
\begin{equation}
\widetilde{\mathcal{L}}^2  \left( \frac{\partial}{\partial t} - \nu \widetilde{\mathcal{L}}^2 \right) 
u_r = 0 \label{nfhs}
\end{equation}

%\subsubsection{Boundary conditions}
%%%%%%%%%%%%%%%%%%%%%%%%%%%%%%%%%%%%

\noindent Since we are interested in the linear stability of the interface located at $ r = R + \zeta$ with $\zeta \ll R$, we Taylor expand the fields and their radial derivatives around $r = R$ and retain only the lowest-order terms,
which are evaluated at $r=R$.
%It is then sufficient to compute the fields and their radial derivatives at $r = R$.
The kinematic condition (\ref{kincondg}) becomes 
\begin{equation}
\frac{\partial}{\partial t} \zeta = u_r|_{r=R} \label{kincondform}
\end{equation}

\noindent We now wish to apply the stress balance equations at $r=R$.
The components of the stress tensor which we will need are

\begin{subequations}\begin{align}
\Pi_{r \theta} & = \eta \left( \frac{1}{r} \frac{\partial u_r}{\partial \theta} 
+ r \frac{\partial}{\partial r} \left( \frac{u_\theta}{r} \right)  \right)\\
\Pi_{r \phi} &= \eta \left( \frac{1}{r \sin \theta} \frac{\partial u_r}{\partial \phi}
+r \frac{\partial}{\partial r} 
\left( \frac{u_\phi}{r} \right)  \right)\\
\Pi_{rr} &= 2\frac{\partial u_r}{\partial r}
\end{align}\label{pitc}\end{subequations}

\noindent We have from \eqref{tstb} that the tangential stress components must vanish at $r=R$, leading to:
\begin{equation}
%\Pi_{r \theta}  = \Pi_{r \phi}  = 0
\Pi_{r \theta}|_{r=R}  = \Pi_{r \phi}|_{r=R} = 0
 \label{tstb2}
\end{equation}

\noindent Taking the horizontal divergence of 
$\Pi_{r\theta}\, \hat{e}_\theta + \Pi_{r\phi}\,\hat{e}_\phi$ leads to
    \begin{align}
0 &=  \left[\nabla_H \cdot (\Pi_{r\theta}\, \hat{e}_\theta + \Pi_{r\phi}\,\hat{e}_\phi ) \right]_{r=R}\nonumber\\
&=\eta\left[\nabla_H\cdot \left(
\frac{1}{r} \frac{\partial u_r}{\partial \theta}\,\hat{e}_\theta + 
\frac{1}{r \sin \theta} \frac{\partial u_r}{\partial \phi}\,\hat{e}_\phi\right)
+\nabla_H\cdot \left( 
r \frac{\partial}{\partial r} \left(\frac{u_\theta}{r}\right) \hat{e}_\theta + 
r \frac{\partial}{\partial r}  \left(\frac{u_\phi}{r}\right)\hat{e}_\phi \right)
\right]_{r=R} \nonumber\\
&=\eta\left[\nabla_H\cdot \nabla_H u_r
+ \frac{\partial}{\partial r}\nabla_H\cdot  \left(u_\theta\hat{e}_\theta+u_\phi\hat{e}_\phi\right)\right]_{r=R}=\eta\left[\left(\nabla_H^2  -\frac{\partial}{\partial r}\frac{1}{r^2}\frac{\partial}{\partial r}r^2\right)u_r\right]_{r=R}
\label{tbc}
   \end{align}
\noindent which is the form of the tangential stress continuity equation that we impose. 

We now wish to linearize the normal stress balance equation:
\begin{equation}
-\left[   \mathbf{n} \cdot \mathbf{\Pi}\cdot \mathbf{n} \right]_{r=R+\zeta}=\left[\mathbf{n} \cdot
\left(P \mathbf{I_d} - \eta \left[ \nabla \bU + \left( \nabla \bU \right)^T \right]\right)\cdot \mathbf{n} \right]_{r=R+\zeta}
= \sigma \left( \frac{1}{R_1} + \frac{1}{R_2} \right)
\label{nsbr}
\end{equation}
The right-hand side of \eqref{nsbr} is $\sigma$ times the curvature of a 
deformed interface and can be shown~\citep[\S275]{Lamb1932} to be, 
up to first order in $\zeta$, 
\begin{equation}
\sigma\left(\frac{1}{R_1} + \frac{1}{R_2} \right)_{r= R + \zeta} \approx  \frac{2\sigma}{R} - \left( \frac{2}{R^2} + \nabla_H^2 \right) \zeta
%\left( \frac{2}{r} + \frac{\ell(\ell+1)}{r^2} \zeta \right)_{r= R + \zeta}
\label{lamb}
\end{equation}
For the left-hand side of \eqref{nsbr}, we use \eqref{BF} to expand the pressure as
\begin{align}
\left(\bar{P}+p \right)_{r=R+\zeta} 
%&
\approx
\left(\bar{P}+p \right)_{r=R}+\left(\frac{\partial\bar{P}}{\partial r}\right)_{r=R}\zeta 
= 2\frac{\sigma}{R}+ p|_{r=R}-\rho G(R,t) \zeta 
\end{align}
Adding the term resulting from the viscosity leads to 
    \begin{align}
-\mathbf{n}\cdot\mathbf{\Pi}\cdot\mathbf{n}|_{r=R+\zeta} \approx 
& 2\frac{\sigma}{R}+ p|_{r=R}-\rho G(R,t) \zeta 
-2\eta\left(\frac{\partial u_r}{\partial r}\right)_{r=R}
\label{lhsnpjc}
\end{align}

\noindent Setting \eqref{lamb} equal to 
\eqref{lhsnpjc} leads to the desired linearized form:

\begin{equation}
p|_{r=R} - \rho G(R,t) \zeta - 2\eta \left(\frac{\partial u_r}{\partial r}\right)_{r=R} = - \sigma \left( \frac{2}{R^2} + \nabla_H^2 \right) \zeta
\label{nst} 
\end{equation}
%LST changed sign of G

\noindent It will be useful to take the horizontal Laplacian of \eqref{nst}:

\begin{equation}
 \nabla_H^2 p|_{r=R} = 2\eta \nabla_H^2 \frac{\partial}{\partial r} u_r |_{r=R} + \rho G(R,t)\nabla_H^2 \zeta - \sigma \nabla_H^2 \left( \frac{2}{R^2} + \nabla_H^2 \right)\zeta %\frac{(l-1)(l+2)}{R^2} \nabla_H^2
\label{thdofnc}
\end{equation}
%LST changed sign of G

%\subsection{Pressure Jump Condition}
%%%%%%%%%%%%%%%%%%%%%%%%%%%%%%%%%%%%%

\noindent We can derive another expression for  $\nabla_H^2 p|_{r=R}$ by taking the horizontal divergence of (\ref{fhs}):

\begin{equation}
    \nabla^2_H p = \frac{1}{r^2}\left(\rho \frac{\partial}{\partial t} - \eta \nabla^2\right) \frac{\partial }{\partial r} (r^2 u_r)  
\label{hdof1}
\end{equation}

\noindent Setting equal the right-hand sides of equations \eqref{hdof1} and \eqref{thdofnc}, we obtain the \emph{pressure jump condition}

\begin{equation}
%    \begin{aligned}
\left[ \frac{1}{r^2}\left(\rho \frac{\partial}{\partial t} - \eta \nabla^2\right) \frac{\partial }{\partial r} (r^2 u_r)    -  2\eta \nabla_H^2 \frac{\partial}{\partial r} u_r \right]_{r=R} = \rho G(R,t)\nabla_H^2 \zeta - \sigma \nabla_H^2 \left( \frac{2}{R^2} + \nabla_H^2 \right)\zeta \\ %\frac{(\ell-1)(\ell+2)}{R^2} \nabla_H^2
%    \end{aligned}
\label{pjcR}
\end{equation}
%LST changed sign of G\\
%
This is the only equation in which the parametrical external forcing appears explicitly; note that only the value $G(r=R,t)$ on the sphere is relevant. 
We now have a set of linear equations \eqref{nfhs}, \eqref{kincondform}, \eqref{tbc} and \eqref{pjcR} involving only $u_r(r,\theta,\phi,t)$ and $\zeta(\theta,\phi,t)$, which reduce to those for the planar case
\citep{KT1994} in the limit of $R\rightarrow\infty$.

%%%%%%%%%%%%%%%%%%%%%%%%%%%%%%%%%%%%%%%%%%%%%%%%%%%%%%%%%%%%%%%%%%%%%%%%
\section{Solution to linear stability problem}
\subsection{Spherical harmonic Decomposition} 
%%%%%%%%%%%%%%%%%%%%%%%%%%%%%%%%%%%%%%%%%%%%%%%%%%%%%%%%%%%%%%%%%%%%%%%%

The equations simplify somewhat when we use 
the poloidal-toroidal decomposition
\begin{equation}
\bu = \curl (f_T \vec{e}_r) + \curl \curl (f \vec{e}_r)
\end{equation}
The radial velocity component $u_r$ depends only on the poloidal field $f$ and is given by
\begin{equation}
u_r (r,\theta,\phi,t) = -\nabla_H^2 f (r,\theta,\phi,t) 
\label{urf}\end{equation}
Using 
\begin{equation}
\underbrace{\left(\frac{1}{r^3}\frac{\partial}{\partial r} r^2\frac{\partial}{\partial r}r  + \nabla^2_H \right)}_{\widetilde{\mathcal{L}}^2}\nabla^2_H = 
\nabla^2_H\underbrace{\left(\frac{\partial^2}{\partial r^2}  + \nabla^2_H\right)}_{\mathcal{L}^2}
\end{equation}
we express \eqref{nfhs} in terms of the poloidal field
\begin{equation}
\nabla_H^2 \left( \frac{\partial}{\partial t} - \nu \mathcal{L}^2 \right) \mathcal{L}^2 f = 0 
\label{nfhspol}\end{equation}
Functions are expanded in series of spherical harmonics $Y_\ell^m(\theta,\phi) = P_\ell^m(\cos \theta) e^{i m \phi}$ satisfiying 
\begin{equation}
 \nabla^2_H Y_\ell^m(\theta,\phi) = - \frac{\ell(\ell +1)}{r^2}Y_\ell^m(\theta,\phi)
\label{basicylm}\end{equation}
We write the deviation of the interface $\zeta(t,\theta,\phi)$ and the scalar function $f$ as
\begin{equation}
\zeta(t,\theta,\phi) = \sum_{\ell=1}^{\infty} \sum_{m=-\ell}^{\ell} \zeta_\ell^m (t) Y_\ell^m(\theta,\phi) \quad \mbox{and} \quad
    f(r,\theta,\phi,t) = \sum_{\ell=1}^{\infty} \sum_{m=-\ell}^{\ell} f_\ell^m(r,t)Y_\ell^m(\theta,\phi)  
\label{expansion}\end{equation}
The equations do not couple different spherical modes $(\ell,m)$, 
allowing us to consider each mode separately. 
The term multiplying $\sigma\nabla_H^2\zeta$ in 
the normal stress equation \eqref{pjcR} becomes 
\begin{equation}
\left(\frac{2}{R^2}+\nabla_H^2\right)\zeta_\ell^m = 
\left(\frac{2}{R^2}-\frac{\ell(\ell+1)}{R^2}\right)\zeta_\ell^m = 
-\frac{(\ell-1)(\ell+2)}{R^2}\zeta_\ell^m
\label{Lamb}\end{equation}
The value $\ell=0$ corresponds to an overall expansion or contraction
of the sphere, which is forbidden by mass conservation and is therefore
excluded from \eqref{expansion}.
The value $\ell=1$ corresponds to a shift of the drop,
rather than a deformation of the interface, so that surface tension
cannot act as a restoring force; 
it is included in this study only when 
the surface tension $\sigma$ is zero and the
constant radial bulk force $g$ is non-zero.

The complete linear problem given by equations \eqref{nfhspol}, \eqref{kincondform}, \eqref{tbc} and \eqref{pjcR} is expressed in terms of $f_\ell^m(r,t)$ and $\zeta_\ell^m(t)$ as:
%
%%%%%%%%%%%%%%%%%%%%%%%%%%%%%%%%%%%%%%%%%%%%%
%\subsection{Boundary condition \label{BClm}}%
%%%%%%%%%%%%%%%%%%%%%%%%%%%%%%%%%%%%%%%%%%%%%
%

\begin{equation}
\left( \frac{\partial}{\partial t} - \nu \mathcal{L}_\ell^2\right) \mathcal{L}_\ell^2 f_\ell^m  = 0 \label{nfhs_ur}
\end{equation}

\begin{equation}
 \frac{\partial}{\partial t}  \zeta_\ell^m = \frac{\ell(\ell +1)}{R^2} f_\ell^m \vert_{r=R} \label{kincondlm}
\end{equation}
 
\begin{equation}
\left( \mathcal{L}^2_\ell - \frac{2}{r}\frac{\partial}{\partial r} \right)f_\ell^m\vert_{r=R} = 0  \label{bc2lm}   
\end{equation}

\begin{align}
\left[ \left(\rho\frac{\partial}{\partial t}\frac{\partial}{\partial r} - \eta \left(\frac{\partial^3}{\partial r^3} +\frac{2}{r}\frac{\partial^2}{\partial r^2} -\ell(\ell+1)\left(\frac{3}{r^2}\frac{\partial }{\partial r} - \frac{4}{r^3} \right)\right)\right)f_\ell^m \right]_{r=R} = \nonumber\\
-  \left(\rho G(R,t) + \sigma \frac{(\ell-1)(\ell+2)}{R^2}\right)\zeta_\ell^m
\label{pjcRlm}\end{align}
%
%CHECK SIGNS\\
%
where we have used \eqref{urf}, \eqref{basicylm} and 
\begin{equation}
  \mathcal{L}^2_\ell \equiv \frac{\partial^2}{\partial r^2}-\frac{\ell(\ell+1)}{r^2}
  \label{Lelldef}\end{equation}
and have divided through by $\ell(\ell+1)/R^2$. 

The value of $m$ does not appear in these equations, 
in much the same way as the Cartesian linear Faraday problem 
depends only on the wavenumber $k$ and not on its orientation.

%\noindent This is the only equation in which the external forcing $G(R,t)$ appears explicitly.

%%%%%%%%%%%%%%%%%%%%%%%%%%%%%%%%%%%%%%%%%%%%%%%%%%%%%%%%%%%%%%%%%%%%%%%%%%%%%%%%%%%%%%%%%%%
\subsection{Floquet Solution}
\label{Floquet Solution}
%%%%%%%%%%%%%%%%%%%%%%%%%%%%%%%%%%%%%%%%%%%%%%%%%%%%%%%%%%%%%%%%%%%%%%%%%%%%%%%%%%%%%%%%%%%

The presence of the time-periodic term in \eqref{pjcRlm} means 
that \eqref{nfhs_ur}, \eqref{kincondlm}, \eqref{bc2lm}, \eqref{pjcRlm}
comprise a Floquet problem. 
To solve it, we follow the procedure of \cite{KT1994}, whereby 
$\zeta_\ell^m$ and $f_\ell^m$ are written in the Floquet form:
\begin{equation}
    \zeta_\ell^m(t) = e^{(\mu + i\alpha)t} \sum_n \zeta_n e^{in\omega t}  \quad   \mbox{and} \quad%\label{flqzlm} 
     f_\ell^m(r,t)  = e^{(\mu + i\alpha)t} \sum_n f_n(r) e^{in\omega t}  
\label{flqdec}
\end{equation}
where $\mu + i\alpha$ is the Floquet exponent and we have omitted the
indices $(\ell,m)$. Substituting the Floquet expansions (\ref{flqdec})
into (\ref{nfhs_ur}) gives, for each temporal frequency $n$
\begin{equation}
\left( \mu+i(n\omega+\alpha) - \nu \mathcal{L}^2_\ell \right) \mathcal{L}^2_\ell f_n  = 0
\end{equation}
or
\begin{equation}
\left( \mathcal{L}^2_\ell - q_n^2 \right) \mathcal{L}^2_\ell f_n(r) = 0 
\label{flqfhs}
\end{equation}
where
\begin{equation}
q_n^2 \equiv \frac{\mu + i(n\omega + \alpha)}{\nu} \label{qn2}
\end{equation}
In order to solve the fourth-order differential equation \eqref{flqfhs}, 
we first solve the homogeneous second-order equation
\begin{equation}
\left( \mathcal{L}^2_\ell - q_n^2  \right) \tilde{f}_n = 0
\label{eqbessel}
\end{equation}
which is a modified spherical Bessel, or Riccati-Bessel, equation~\citep{AS65}. 
The general solutions are of the form 
\begin{equation}
\tilde{f}_n(r) = \tilde{B}_n r^{\frac{1}{2}}J_{\ell+\frac{1}{2}}(iq_n r) + \tilde{D}_n^2 r^{\frac{1}{2}}J_{-(\ell+\frac{1}{2})}(iq_n r)
\end{equation}
where $J_{\ell+\frac{1}{2}}$ is the spherical Bessel function of half-integer order $\ell + \frac{1}{2}$. 

The remaining second-order differential equation is 
\begin{equation}
\mathcal{L}^2_\ell f_n= \tilde{f}_n
\end{equation}
whose solutions are the general solutions of \eqref{flqfhs} and are given by 
\begin{equation}
f_n(r) = A_n r^{\ell+1} + B_n r^{\frac{1}{2}}J_{\ell+\frac{1}{2}}(iq_n r) + C_n r^{-\ell} + D_n r^{\frac{1}{2}}J_{-(\ell+\frac{1}{2})}(iq_n r)
\label{flqsolgen}
\end{equation}
Note that $u_r\sim f_n/r^2$ satisfies \eqref{nfhs}.
Eliminating the solutions in \eqref{flqsolgen} 
which diverge at the centre, we are left with:
\begin{equation}
f_n(r) = A_n r^{\ell+1} + B_n r^{\frac{1}{2}}J_{\ell+\frac{1}{2}}(iq_n r)
\label{flqsolfin}
\end{equation}
The constants $A_n$ and $B_n$ can be related to $\zeta_n$ by using the kinematic condition \eqref{kincondlm} and the tangential stress condition \eqref{bc2lm} which, for  Floquet mode $n$, are 

\begin{align}
(\mu+i(n\omega+\alpha)) &\zeta_n = \frac{\ell(\ell+1)}{R^2} f_n|_{r=R} \label{kincondflq}\\
\left( \frac{\partial^2}{\partial r^2} - \frac{2}{r}\frac{\partial}{\partial r} + \frac{\ell(\ell +1)}{r^2}\right) & f_n\vert_{r=R} = 0  \label{bc2flq}   
%\label{kcbc2}
\end{align}
Appendix \ref{appendix} gives more details on the determination of 
these constants and also presents the solution and boundary conditions 
in the case for which the exterior of the drop is a fluid rather than a vacuum.

There remains the normal stress (pressure jump) condition (\ref{pjcRlm}), 
the only equation which couples temporal Floquet modes for different $n$. 
Writing 
\begin{equation}
a \cos(\omega t)\sum_n \zeta_ne^{in\omega t} = a \frac{e^{i\omega t}+ e^{-i\omega t}}{2} \sum_n\zeta_ne^{in\omega t} = \frac{a}{2}\sum_n(\zeta_{n+1} + \zeta_{n-1}) e^{in\omega t}
\end{equation}
and inserting the Floquet decomposition (\ref{flqdec}) into (\ref{pjcRlm})
leads to

\begin{align}
  \left[ \left(\rho(\mu+i(n\omega + \alpha)) \frac{\partial}{\partial r}\right.\right.
    &\left.\left.- \eta\left(\frac{\partial^3}{\partial r^3} +\frac{2}{r}\frac{\partial^2}{\partial r^2} -\ell(\ell+1)\left(\frac{3}{r^2}\frac{\partial }{\partial r}  -\frac{4}{r^3} \right)\right)\right)f_n \right]_{r=R} &\nonumber\\
&+  \left(\rho g + \sigma \frac{(\ell-1)(\ell+2)}{R^2}\right)\zeta_n 
 = \rho \frac{a}{2}(\zeta_{n+1}+\zeta_{n-1})
 \label{pjcfn}
\end{align}
Using \eqref{derfn} and \eqref{derbf} to express the partial derivatives of
$f_n$ as multiples of $\zeta_n$, \eqref{pjcfn} can be written as an 
eigenvalue problem with eigenvalues $a$ and eigenvectors $\{\zeta_n\}$ 
\begin{align}
\mathcal{A}\zeta &= a\mathcal{B}\zeta \label{eigvalpb}
\end{align}
The usual procedure for a stability analysis is to fix the wavenumber
(here the spherical mode $\ell$) and the forcing amplitude $a$, to
calculate the exponents $\mu + i \alpha$ of the growing solutions and
to select that whose growth rate $\mu(\ell,a)$ is largest. Following
\cite{KT1994}, we instead fix $\mu = 0$ and restrict
consideration to two kinds of growing solutions, harmonic with $\alpha
= 0$ and subharmonic with $\alpha = \frac{\omega}{2}$. We then solve
the problem \eqref{eigvalpb} for the eigenvalues $a$
and eigenvectors $\{\zeta_n\}$ and select the smallest, or
several smallest, real positive eigenvalues $a$.
These give the marginal stability curves
$a(\ell,\mu=0,\alpha=\frac{\omega}{2}$ and $a(\ell,\mu=0,\alpha=0)$
without interpolation.
Our computations require no more than 10 Fourier modes in
representation \eqref{flqdec}.
This method can be used to solve any Floquet problem for which
the overall amplitude of the time-periodic terms can be varied.
The detailed procedure for the solution of
the eigenvalue problem (\ref{eigvalpb}) is described in 
\cite{KT1994,K1996}.

%%%%%%%%%%%%%%%%%%%%%%%%%%%%%%%%%%%%%%%%%%%%%%%%%%%%%%%%%%%%%%%%%%%%%%%%%%%%%%%%%%%%%%%%%%%%%%%%%%%%%%%%%%%%%%%%%%%%%%%%%%%%%%%%%%%%%%%%%%%%%%%%%%%%%%%%%%%%%%%%%%%%%%
\section{Ideal fluid case and non-dimensionalization}
\label{sec:ideal}
%%%%%%%%%%%%%%%%%%%%%%%%%%%%%%%%%%%%%%%%%%%%%%%%%%%%%%%%%%%%%%%%%%%%%%%%%%%%%%%%%%%%%%%%%%%%%%%%%%%%%%%%%%%%%%%%%%%%%%%%%%%%%%%%%%%%%%%%%%%%%%%%%%%%%%%%%%%%%%%%%%%%%%%

For an ideal fluid drop ($\nu = 0$) and for a given spherical harmonic
$Y_\ell^m$, system (\ref{nfhs_ur}) reduces to 

\begin{equation}
 \frac{\partial}{\partial t} \mathcal{L}^2_\ell f(r,t) = 0 
 \label{idfhs}
\end{equation}

\noindent 
We make the customary assumption that
%the vorticity
$\mathcal{L}^2_\ell f(r,t)$
is not only constant, as implied by \eqref{idfhs}, but zero. In this case, 
the solution which does not diverge at the drop centre is of the form
\begin{equation}
f(r,t) = F(t) r^{\ell+1}
\label{idfn}
\end{equation}

\noindent As the tangential stress is purely viscous in origin, only the kinematic condition \eqref{kincondlm} and the pressure jump condition across the interface \eqref{pjcRlm} are applied. Using \eqref{idfn}, 
these are reduced to:

\begin{equation}
\dot{\zeta}(t)  = \frac{\ell(\ell+1)}{R^2} F(t) R^{\ell+1}
 \label{idkincond}
\end{equation}

\begin{equation}
(\ell+1)\dot{F}(t) R^\ell =  -\left(G(R,t)+\frac{\sigma}{\rho} \frac{(\ell-1)(\ell+2)}{R^2} \right) \zeta_\ell^m(t)
\label{idpjclm}
\end{equation}
%CHECK SIGNS

\noindent By differentiating \eqref{idkincond} with respect to time and substituting into \eqref{idpjclm}, we arrive at
\begin{equation}
\ddot{\zeta} =  -\left(g\frac{\ell}{R}+\frac{\sigma}{\rho} \frac{\ell(\ell-1)(\ell+2)}{R^3} -a\frac{\ell}{R}\cos(\omega t)\right) \zeta
\label{beforenondim}\end{equation}
Focusing for the moment on the unforced equation, we define:
\begin{align}
   \omega_0^2 & \equiv \left(g \frac{\ell}{R}+ \frac{\sigma}{\rho} \frac{\ell(\ell-1)(\ell+2)}{R^3}\right)   \label{RayEigFr}
\end{align}
Equation \eqref{RayEigFr} is the spherical analogue of 
the usual dispersion relation for gravity-capillary waves 
in a plane layer of infinite depth, with the modifications 
\begin{subequations}
\begin{align}
gk &\rightarrow g\frac{\ell}{R} \label{gcorr}\\
\frac{\sigma}{\rho}k^3 &\rightarrow \frac{\sigma}{\rho} \frac{\ell(\ell-1)(\ell+2)}{R^3}\label{sigcorr}
\end{align}
\label{matpar}\end{subequations}
The transformation \eqref{gcorr} can be readily understood by the 
fact that the wavelength of a spherical mode $\ell$ is $2\pi R/\ell$.
The transformation \eqref{sigcorr} must be understood in light 
of the fact that an unperturbed sphere, unlike a planar surface, 
has a non-zero curvature term proportional to $2\sigma/R$, 
from which \eqref{sigcorr} is derived as a deviation via \eqref{Lamb}.
In the absence of a bulk force, $g=0$, \eqref{RayEigFr} becomes
the formula of \cite{Rayl1879} or \cite{Lamb1932} for the eigenfrequencies 
of capillary oscillation of a spherical drop 
perturbed by a deformation characterized by spherical wavenumber $\ell$.

Using definitions \eqref{RayEigFr} and
\begin{equation}
  a_0 \equiv \frac{R \omega_0^2}{\ell}
\label{a0def}  \end{equation}
and non-dimensionalizing time via $\hat{t}\equiv t\omega$,
equation \eqref{beforenondim} can be converted to the Mathieu equation
\begin{equation}
  \frac{d^2\zeta}{d\hat{t}^2} = -\left(\frac{\omega_0}{\omega}\right)^2
  \left(1 - \frac{a}{a_0}\cos\hat{t}\right)\zeta %= 0
\label{basiceq1}  \end{equation}
combining the multiple parameters 
$g$, $R$, $\sigma$, $\rho$, $a$, $\omega$ and $\ell$
into only two, $\omega/\omega_0$ and $a/a_0$.
The stability regions of \eqref{basiceq1} are bounded by tongues which
intersect the $a=0$ axis at
\begin{equation}
\omega = \frac{2}{n}\omega_0
\label{basiceq2}\end{equation}
Thus the inviscid spherical Faraday linear stability problem
reduces to the Mathieu equation, as it does in the planar case 
\citep{BU1954}.
In a Faraday wave experiment or numerical simulation,
$\ell$, unlike the other parameters, is not known {\it a priori}. 
Instead, in light of \eqref{RayEigFr}, equation \eqref{basiceq2}
is a cubic equation which determines $\ell$ given the other parameters. 
Since $\omega_0$ and $a_0$ are functions of $\ell$, 
both of the variables $\omega/\omega_0$ and $a/a_0$ contain
the unknown $\ell$ and so cannot be interpreted as
simple non-dimensionalizations of $\omega$ and $a$. 
(For the purely gravitational case, $a_0=g$ is independent of $\ell$.) 

\begin{figure}
\vspace*{0.2cm}
\includegraphics[width=\textwidth]{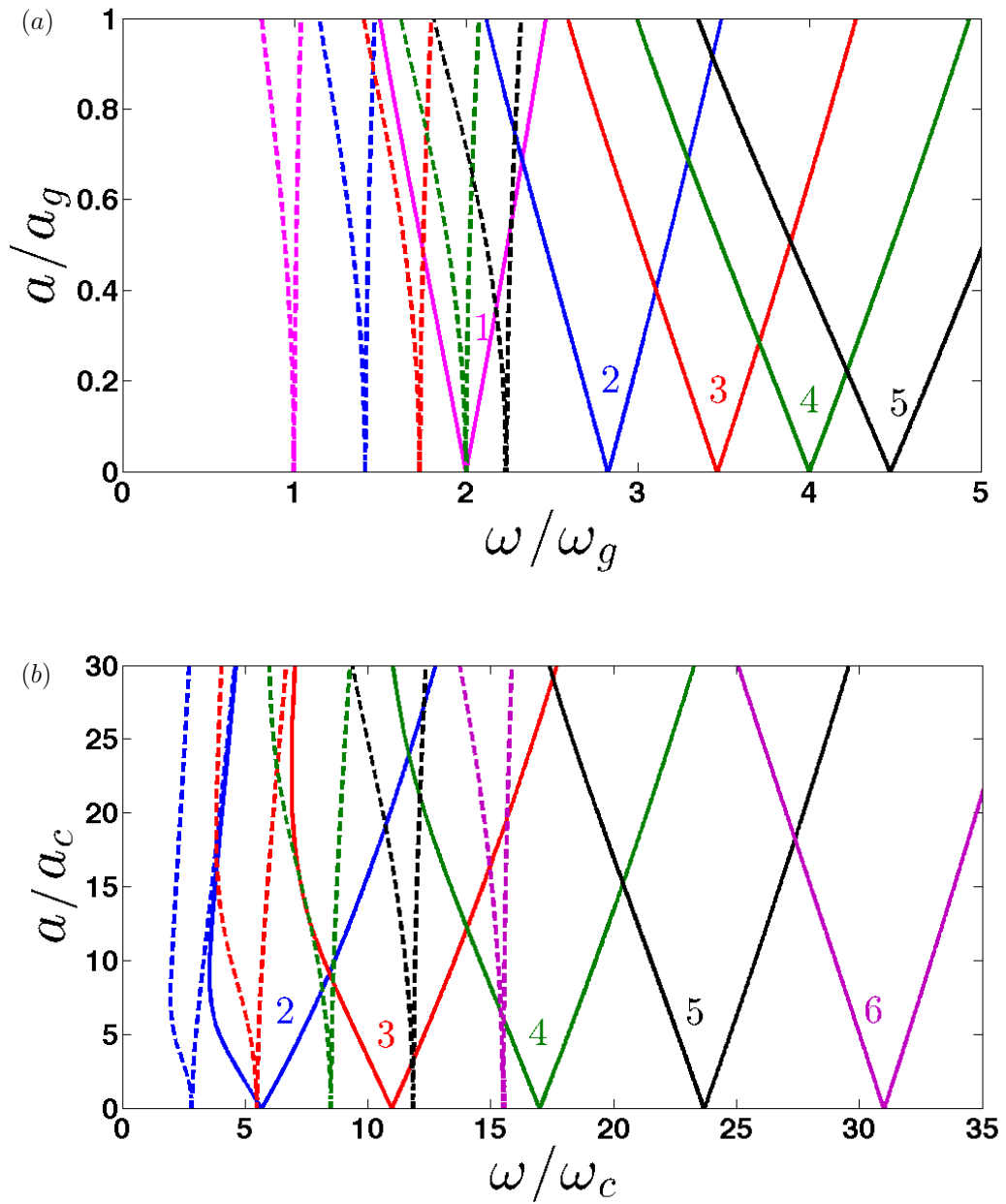}
  \caption{Instability tongues of an inviscid fluid drop due to oscillatory
    forcing with amplitude $a$ and angular frequency $\omega$.  Solid
    curves bound subharmonic tongues and dashed curves bound harmonic
    tongues.  (a) Tongues
    corresponding to gravitational instability with spherical 
    wavenumbers $\ell=1$, 2, 3, 4 originate at $\omega/\omega_g =
    2\sqrt{\ell}/n$.
(b) Tongues corresponding to capillary instability with
    spherical wavenumbers $\ell=2$, 3, 4, 5 originate at
    $\omega/\omega_c =2\sqrt{\ell(\ell-1)(\ell+2)}/n$, with $n$ odd
    (even) for subharmonic (harmonic) tongues.}
	\label{fig:mathitong}
\end{figure}
\begin{figure}
\begin{center}
	\includegraphics[width=\textwidth]{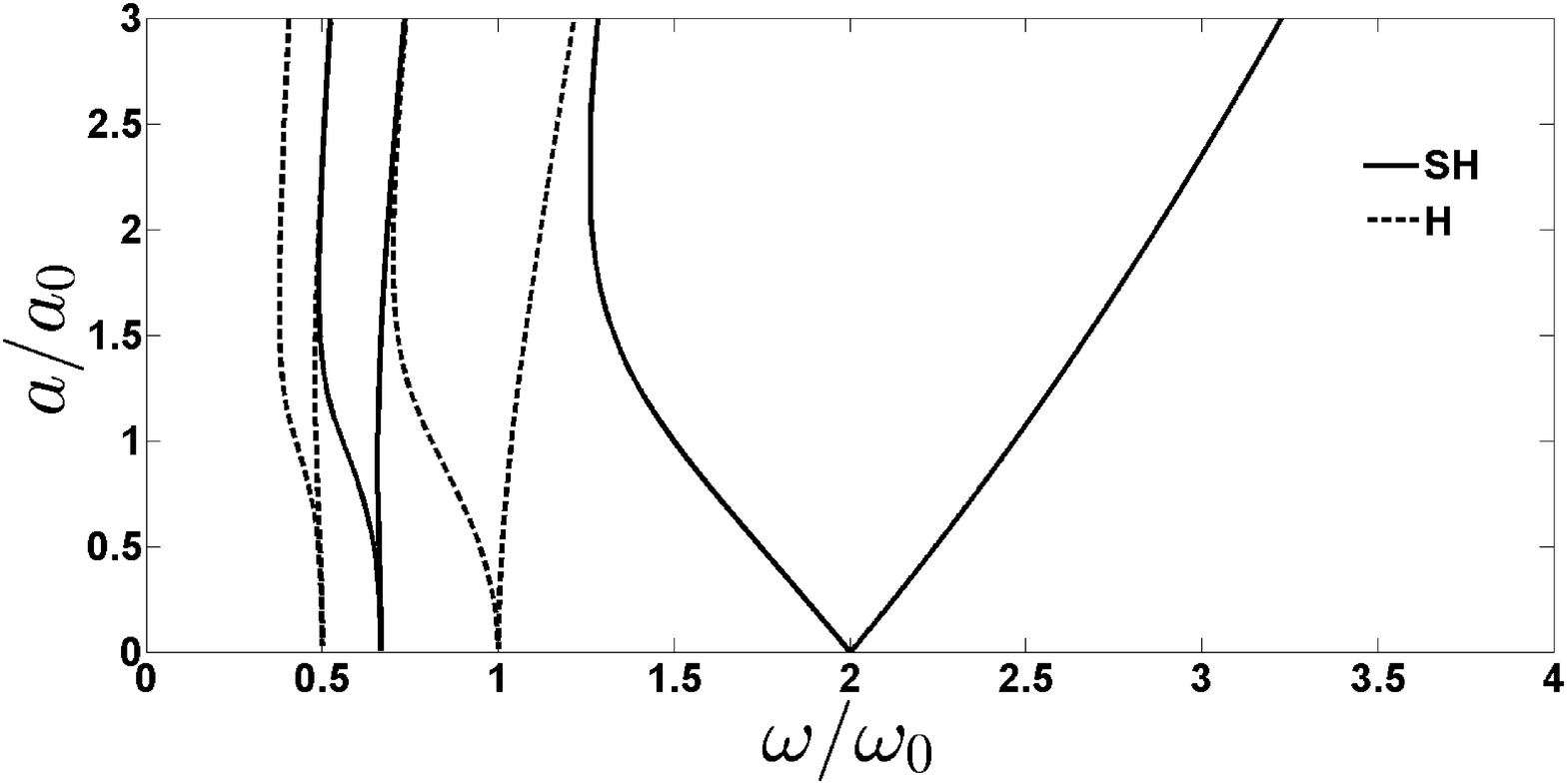}
	\caption{Same data as in figure \ref{fig:mathitong}, but
          scaled by $\omega_0$ and $a_0$.  The tongues for the
          gravitational and the capillary cases and for all values of
          $\ell$ all collapse onto a single set of tongues.  Solid
          curves bound subharmonic tongues and dashed curves bound
          harmonic tongue.}
	\label{fig:Mat}
\end{center}
\end{figure}

It is useful to examine \eqref{RayEigFr} and \eqref{a0def} in the
two limits of gravity and capillary waves.
We first define non-dimensional angular frequencies and oscillation amplitudes
which do not depend on $\ell$:
\begin{eqnarray}
\omega_g^2 \equiv \frac{g}{R} &&\qquad a_g \equiv R\omega_g^2 = g   \\
\omega_c^2 \equiv \frac{\sigma}{\rho R^3}&& \qquad a_c \equiv R\omega_c^2
\end{eqnarray}
so that 
\begin{eqnarray}
  \omega_0^2 &=& \omega_g^2 \;\ell + \omega_c^2 \;\ell(\ell-1)(\ell+2) \\
  a_0 &=& a_g + a_c (\ell-1)(\ell+2)
  \end{eqnarray}
The Bond number measuring the relative importance of the two forces can be written as:
\begin{equation}
  Bo \equiv \frac{\rho g R^2}{\sigma} = \frac{\omega_g^2}{\omega_c^2}
\end{equation}

In the gravity-dominated regime (large $Bo$), we write \eqref{basiceq2} as 
\begin{equation}
  \omega^2 =   \left(\frac{2}{n}\right)^2 \omega_g^2 \; \left[\ell + \frac{1}{Bo}\ell (\ell-1)(\ell+2)\right]
\end{equation}
The stability boundaries for $1/Bo=0$ 
are given in figure \ref{fig:mathitong}(a).
The subharmonic and harmonic tongues originate at
$\omega/\omega_g = 2\sqrt{\ell}$ and $\omega/\omega_g = \sqrt{\ell}$.

In the capillary-dominated regime (small $Bo$)
it is more appropriate to write \eqref{basiceq2} as 
\begin{equation}
  \omega^2 = \left(\frac{2}{n}\right)^2 \omega_c^2 \left[\ell (\ell-1)(\ell+2)
  + Bo \; \ell\right]
\end{equation}
The stability boundaries for $Bo=0$ are given in figure \ref{fig:mathitong}(b)
in terms of $\omega/\omega_c$ and $a/a_c$.
These consist of families of tongues, which originate on the $a=0$
axis at $\omega/\omega_c=2\sqrt{\ell(\ell-1)(\ell+2)}/n$ for
$\ell=2$, 3, 4, 5 and for $n=1$, 2, ...
within which the drop has one of the spatial forms corresponding
to the spherical wavenumber $\ell$ and oscillates like $e^{i n \omega t/2}$. 
The solid curves bound the first subharmonic instability tongues, which
orginate on the $a=0$ axis at 
$\omega/\omega_c = 2\sqrt{\ell(\ell-1)(\ell+2)}$ for $\ell=2$, 3, 4, 5,
within which the drop oscillates like $e^{i\omega t/2}$.
The dashed curves bound the first harmonic tongues, originating at 
$\omega/\omega_c = \sqrt{\ell(\ell-1)(\ell+2)}$
describing a response like $e^{i\omega t}$.
Tongues for higher $n$ are located at still lower values of $\omega$. 

The curves in figure \ref{fig:mathitong}
for different $\ell$, $g$, $\sigma/\rho$, $R$ all collapse onto a single set
of tongues when they are plotted in terms of
$\omega/\omega_0$ and $a/a_0$.
This is shown in figure \ref{fig:Mat},
in which the various tongues correspond to the temporal harmonic index $n$.
In order to use figure \ref{fig:Mat} to determine
whether the drop is stable against perturbations 
with spherical wavenumber $\ell$ when 
a radial force with amplitude $a$ and angular frequency $\omega$ is applied, 
the following procedure must be used. 

For each $\ell$, formulas \eqref{RayEigFr} and \eqref{a0def} are used
to determine the values of $(\omega_0,a_0)$,
If $(\omega/\omega_0,a/a_0)$ is inside one of the instability tongues
(usually, but not always, that corresponding to an $\omega/2$ response
with $n=1$) 
then the drop is unstable to perturbations of that $\ell$. 
The drop is stable if $(\omega/\omega_0, a/a_0)$ lies outside the tongues
for all $\ell$ and all $n$. 
This is the procedure described by \cite{BU1954}
and carried out by \cite{Batson} in a cylindrical geometry.

Because $(\omega/\omega_0,a/a_0)$ depends on the unknown $\ell$, 
an experimental or numerical choice of parameters cannot
immediately be assigned to a point in figure \ref{fig:Mat},
rendering its interpretation somewhat more obscure.
It is perhaps because of this that investigations of the Faraday
instability are often presented in dimensional terms.
Figures like \ref{fig:mathitong}(a) and (b),
in which the two axes are non-dimensional quantities defined in
terms of known parameters, 
represent a good compromise between universality and ease of use.
This treatment can also be applied to non-spherical geometries in which the
unperturbed surface is flat and the depth is finite.

%%%%%%%%%%%%%%%%%%%%%%%%%%%%%%%%%%%%%%%%%%%%%%%%%%%%%%%%%%%%%%%%%%%%%%%%%%%%%%%%%%%%%%%%%%%%%%%%%%%%%%%%%%%%%%%%%%%%%%%%%%%%%%%%%%%%%%%%%%%%%%%%%%%%%%%%%%%%%%%%%%%%%%
\section{Viscous fluids and scaling laws}
%%%%%%%%%%%%%%%%%%%%%%%%%%%%%%%%%%%%%%%%%%%%%%%%%%%%%%%%%%%%%%%%%%%%%%%%%%%%%%%%%%%%%%%%%%%%%%%%%%%%%%%%%%%%%%%%%%%%%%%%%%%%%%%%%%%%%%%%%%%%%%%%%%%%%%%%%%%%%%%%%%%%%%%

We now return to the viscous case.  For a viscous fluid, it is not
possible to reduce the governing equations to a Mathieu equation even
with the addition of a damping term \citep{KT1994}.  As described in
section \ref{Floquet Solution}, the governing equations and boundary
conditions are reduced to an eigenvalue problem, whose solution gives
the critical oscillation amplitude $a$ as an eigenvalue.

\begin{figure}
\includegraphics[width=\textwidth,clip]{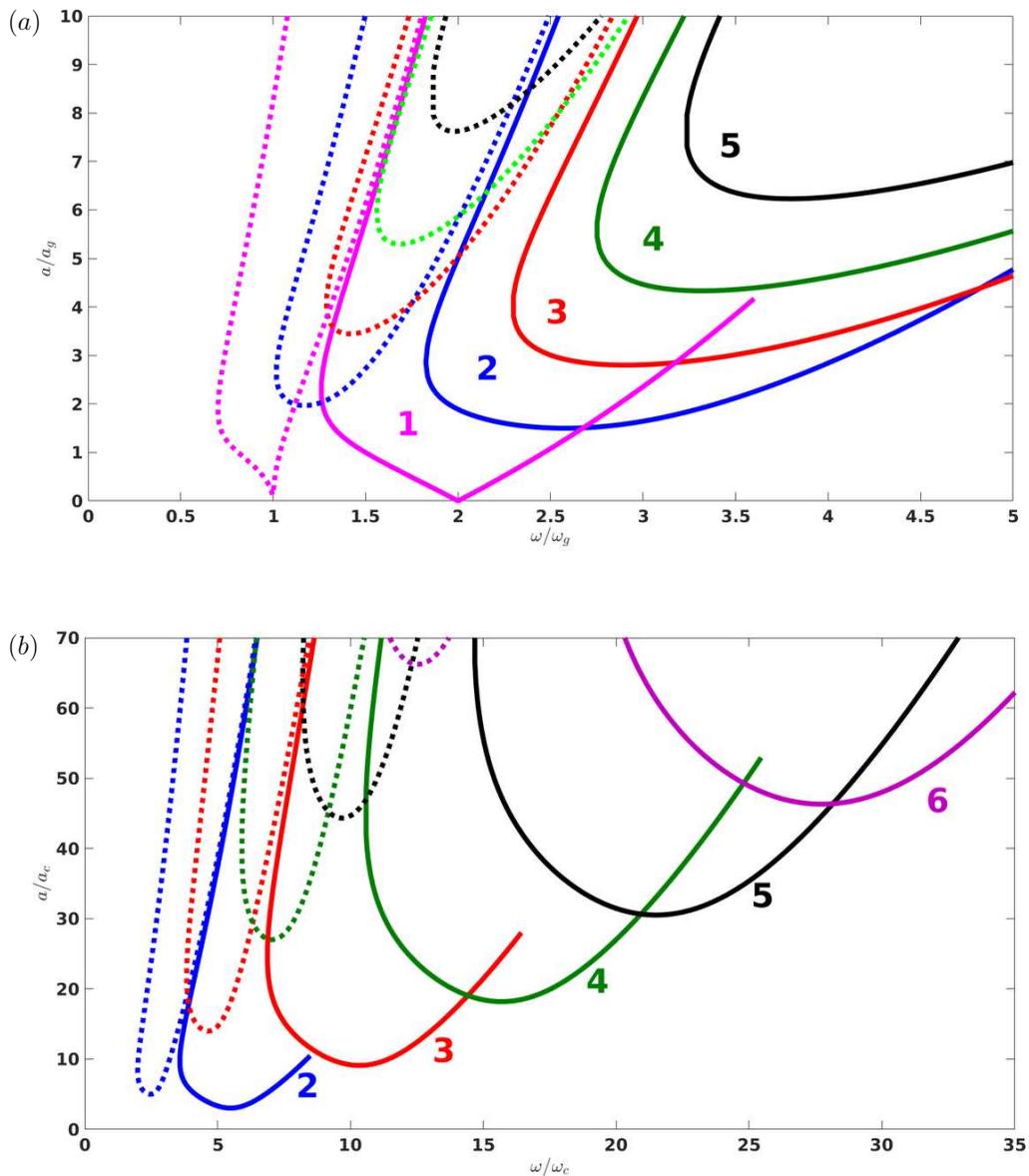}
\caption{Instability tongues due to oscillatory forcing of amplitude
  $a$ and angular frequency $\omega$ for a viscous fluid drop.  Solid
  curves bound subharmonic tongues and dashed curves bound harmonic
  tongues.  (a) Tongues corresponding to gravitational instability
  with viscosity $\nu/\sqrt{g R^3}=0.08$.  Spherical wavenumbers are
  $\ell=1$ (magenta), 2 (blue), 3 (red), 4 (green), 5 (black).
  (b) Tongues correspond to capillary instability
  with viscosity $\nu/\sqrt{\sigma R/\rho}=0.08$.  Spherical
  wavenumbers are $\ell=2$ (blue), 3 (red), 4 (green), 5 (black), 6 (purple).}
	\label{fig:mathvtong}
\end{figure}

Figure \ref{fig:mathvtong} displays the regions of instability
of a viscous drop using the same conventions as we did
for the ideal fluid case, i.e. treating capillary and gravitational
instability separately and plotting the stability boundaries in units
of $\omega_c$, $\omega_g$, $a_c$, $a_g$.
Viscosity smoothes the
instability tongues and displaces the critical forcing amplitude
towards higher values, with a displacement which increases with $\ell$.
The viscosity used in figure \ref{fig:mathvtong} is 
$\nu/\sqrt{\sigma R/\rho}=0.08$ and $\nu/\sqrt{g R^3}=0.08$
the Ohnesorge number and the inverse square root of the Galileo number
for the capillary and the gravitational cases, respectively.
This value is chosen to be high enough to show the important qualitative effect
of viscosity, but low enough to permit the first few tongues to
be shown on a single diagram. It can be seen that the effect of viscosity
is greater on tongues with higher $\ell$; this important point will
be discussed below.
For low enough frequency, i.e. $\omega/\omega_c\lesssim 1.2$ in figure \ref{fig:mathvtong}(a)
and $\omega/\omega_g\lesssim 4$ in figure \ref{fig:mathvtong}(b), it can be seen that 
the instability is harmonic rather than subharmonic, as discussed by \cite{K1996}. 

\begin{figure}
\includegraphics[height=0.8\textheight]{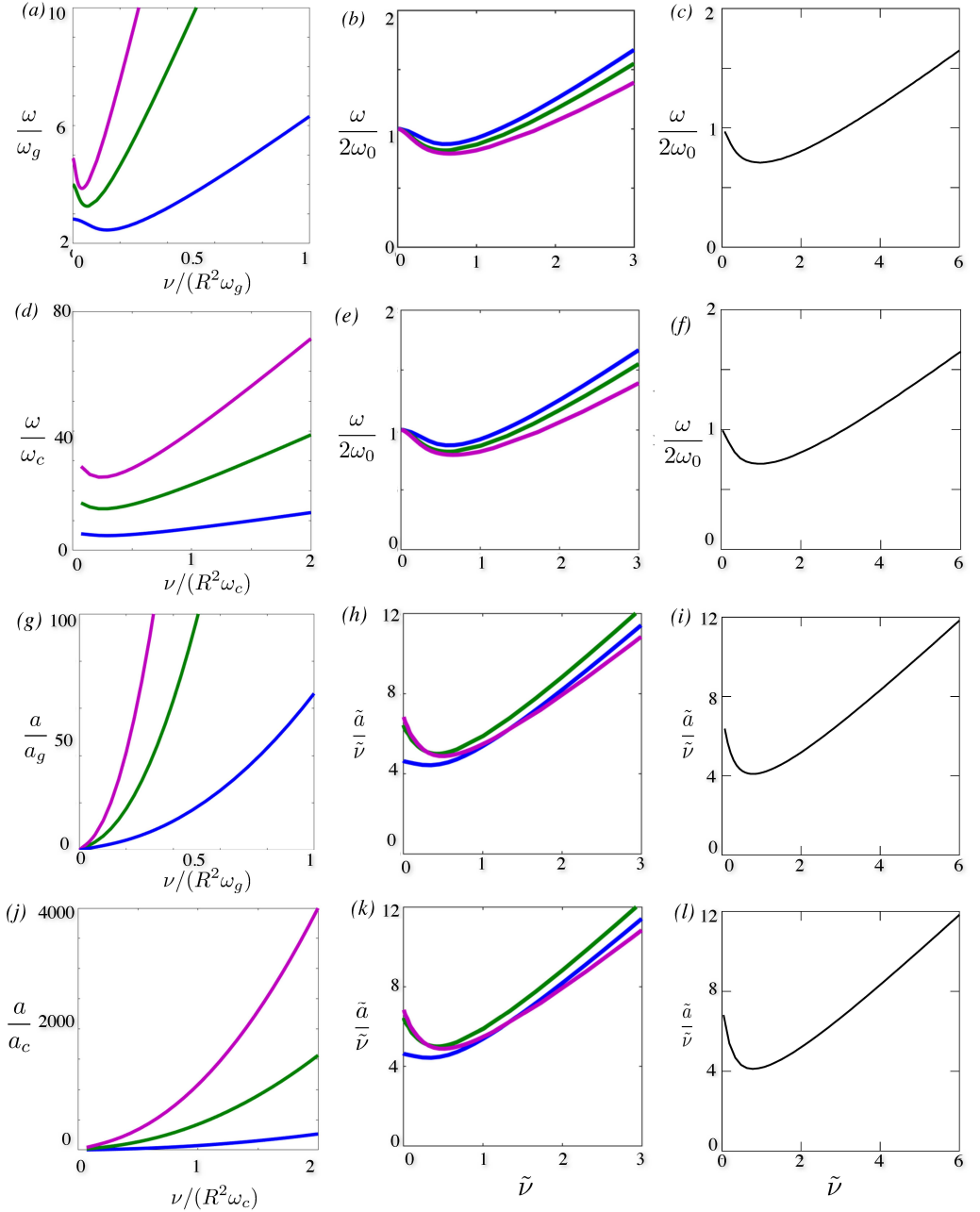}
\caption{Viscosity dependence of $\omega$ (rows 1 and 2) and $a$
  (rows 3 and 4) at threshold. Gravitational (rows 1 and 3) and
  capillary (rows 2 and 4) Faraday instability shown for 
  $\ell=2$ (blue), 4 (green), and 6 (purple).
    Leftmost column (a,d,g,j) $\omega$ and $a$ non-dimensionalized in terms of
    experimentally imposed quantities $\omega_g$, $\omega_c$, $a_g$ and
    $a_c$. Non-dimensionalization of middle column (b,e,h,k) uses
     (\ref{eq:nugood}a) and (\ref{eq:agood}a).
     %    $R/\sqrt{\ell(\ell+1)}$ as length scale and $(\omega_0)^{-1}$ as time scale.
     It can be seen that $a\sim\nu$ for $\nu$ small.
    Right column (c,f,i,l) shows analogous quantities for the planar infinite-depth case, calculated using 
      (\ref{eq:nugood}b) and (\ref{eq:agood}b); the scaling is seen to be
      exact in this case.}
	\label{fig:viscdep}
\end{figure}

Figure \ref{fig:viscdep} shows the variation with viscosity of the Faraday 
threshold, more specifically of the coordinates $(\omega,a)$ of the
minimum of the primary subharmonic tongue, for $\ell=2,4,6$
for the capillary and gravitational Faraday instability.
The first column shows this dependence using non-dimensional
quantities that are independent of $\ell$. (We explained the motivation 
for such a choice in section \ref{sec:ideal}, namely that
$\ell$ is not known a priori in an experiment or full numerical simulation.)
The second column shows the non-dimensionalization that best 
fits the data for all values of $\ell$. The appropriate choice for
the non-dimensionalization of viscosity is 
\refstepcounter{equation}\label{eq:nugood}
\begin{equation}
\tilde{\nu}\equiv \frac{\nu\ell(\ell+1)}{R^2\omega_0} 
 \qquad\mbox{or}\qquad
\tilde{\nu}\equiv \frac{\nu k^2}{\omega_0}
\tag{\theequation a,b}\label{eq:nugoodab}
\end{equation}
in the spherical or planar geometries, respectively, based on the wavelength.
See \cite{BEMJ1995}, who studied the influence of viscosity on Faraday waves.
The choice of \eqref{eq:nugood} is guided by comparing the viscous and oscillatory
terms in \eqref{pjcfn} and corroborated by the fact that the curves in the
second column have only a weak dependence on $\ell$ and on $\tilde{\nu}$. 
We recall that the ratio of the horizontal (angular)
wavelength $2\pi R/\ell$ to the depth (radius) $R$ goes to zero
as $\ell$ increases, and the curvature of the sphere is less
manifested over a horizontal wavelength.
With increasing $\ell$, 
the curves for the spherical case can be seen to approach the 
corresponding quantities in the planar infinite-depth case,
shown in the third column, despite the fact that we are
far from the large $\ell$ limit.

Figures \ref{fig:viscdep}(a)-(f) show that the frequency $\omega$
which favours waves (corresponding to the bottom
of the tongue) is a non-monotonic function
of $\nu$, for which an explanation is proposed by \cite{KT1994}.
At lower viscosities, the flow is assumed to be irrotational,
as in \eqref{idfn}, and equation \eqref{RayEigFr} is modified
merely by subtracting a term proportional to $\nu^2$.
This leads to a decrease in the critical $\omega$ from $2\omega_0$.
At higher viscosities, it is assumed that the response time $4\pi/\omega$
approaches the viscous time scale, here $O(\ell(\ell+1)R^2/\nu)$, leading to
an increase in $\omega$ with $\nu$ when the other parameters are fixed.
Experimental values for $\tilde{nu}$ are, however, rarely greater than one.

Concerning the oscillation amplitude $a$, 
we find that the appropriate choice for non-dimensionalization is
\refstepcounter{equation}\label{eq:agood}
\begin{equation}
  \tilde{a}\equiv \frac{a \ell}{R\omega_0^2}
\qquad \mbox{or}\qquad
  \tilde{a}\equiv \frac{a k}{\omega_0^2}
  \tag{\theequation a,b} \label{eq:agoodab}
\end{equation}
This non-dimensionalization causes all six curves of figure
\ref{fig:viscdep}(g) and (j) to collapse.
For small viscosities, $\tilde{a}$ increases linearly with $\tilde{\nu}$;
see \cite{BEMJ1995}.
For this reason we plot 
\refstepcounter{equation}
\begin{equation}
\frac{\tilde{a}}{\tilde{\nu}}=  \frac{a \ell}{R\omega_0^2}
    \left(\frac{\nu\ell(\ell+1)}{R^2\omega_0}\right)^{-1}
  =\frac{aR}{\nu\omega_0(\ell+1)}
\qquad \mbox{or}\qquad
\frac{\tilde{a}}{\tilde{\nu}}=\frac{a}{\nu\omega_0 k}
\tag{\theequation a,b}\end{equation}
in figures \ref{fig:viscdep}(h,i,k,l).
The form of this curve for higher viscosities shows that 
$\tilde{a}$ contains terms of higher order in $\tilde{\nu}$,
as demonstrated by \cite{VKM2001}.
  
The viscosity dependence of $\omega$ and $a$,
once they are non-dimensionalized,
are practically identical
for the capillary and gravitational cases.
The difference between these, as well as the dependence on $\ell$,
is taken into account exclusively via $\omega_0(\ell)$,
just as it is for the inviscid problem.

\section{Eigenmodes}

Thus far, we have not discussed the spatial form of the eigenmodes
on the interface, beyond stating that they are spherical harmonics.
Visualizations of the spherical harmonics, while common,
are inadequate or incomplete for depicting the
behaviour of the interface in this problem for a number of reasons. 
First, as stated in the introduction, the linear stability
problem is degenerate: the $2\ell+1$ spherical harmonics
which share the same $\ell$ all have the same linear growth rates
and threshold.
Second, for $\ell\geq 4$, the patterns actually realized in
experiments or numerical simulations, which are determined by the
nonlinear terms, are not individual spherical harmonics, but
particular combinations of them.
Finally, in a Floquet problem, the motion of the interface
is time dependent.

\begin{figure}
\includegraphics[width=\textwidth]{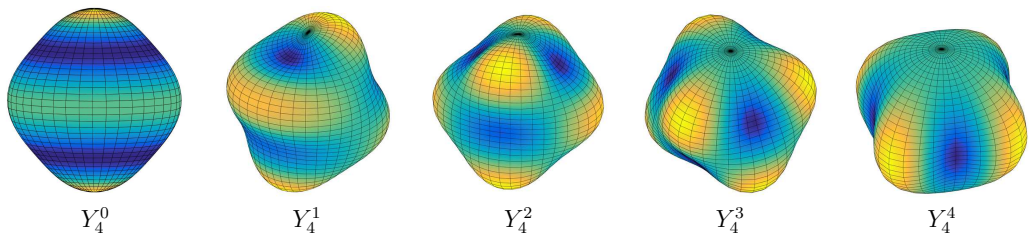}
\caption{Spherical harmonics for $\ell=4$.}
\label{fig:pics1}
\end{figure}

\begin{figure}
\includegraphics[width=\textwidth]{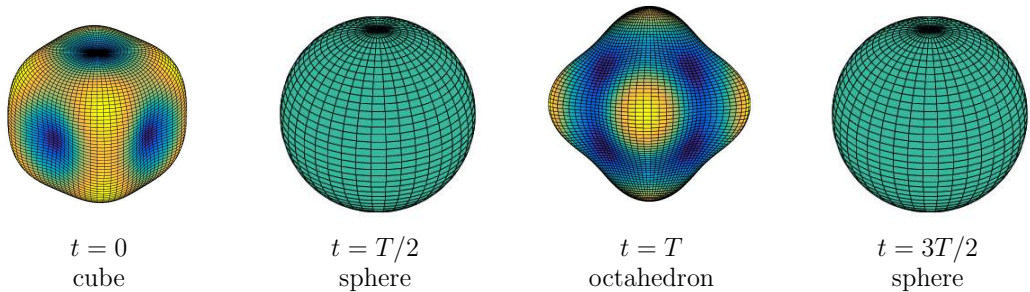}
\caption{Subharmonic $\ell=4$ standing wave pattern oscillates
  in time between cubic and octahedral shapes.}
\label{fig:pics2}
\end{figure}

Figure \ref{fig:pics1} shows the spherical harmonics for $\ell=4$. 
Spherical harmonics can be 
classified as zonal ($m=0$, independent of $\phi$, nodal lines which
are circles of constant latitude), sectoral ($m=\pm\ell$, independent
of $\theta$, nodal lines which are circles of constant longitude), or
tesseral ($m\neq 0,\pm\ell$, checkered).  
Figure \ref{fig:pics2} depicts the pattern that is realized
in numerical simulations for $\ell=4$. The pattern oscillates between a
cube and an octahedron and is a combination of $Y_4^0$ and $Y_4^4$. 
This pattern is the result of nonlinear selection;
at the linear level, many other patterns could be realized.
Our companion paper is devoted to a comprehensive description 
of the motion of the interface and of the velocity field
for $\ell$ between 1 and 6.


\section{Discussion}

We have considered a configuration similar to the classic problem of
freely oscillating liquid drops, that of a viscous drop under the
influence of a time-periodic radial bulk force and of surface tension.
Here, we have carried out a linear stability analysis, while in a
companion paper \cite{Ali2}, we describe the results of a full
numerical simulation.  We believe both of these investigations to be
the first of their kind.

Our investigation relies on a variety of mathematical and
computational tools.  We have solved the linear stability problem by
adapting to spherical coordinates the Floquet method of \cite{KT1994}.
The solution method uses a spherical harmonic decomposition in
the angular directions and a Floquet decomposition in time to reduce
the problem to a series of radial equations, whose solutions are
monomials and spherical Bessel functions.
We find that the equations for the inviscid case reduce exactly to the
Mathieu equation, as they do for the planar case \cite{BU1954}, with
merely a reassignment of the parameter definitions.  In contrast, for
the viscous case, there are additional terms specific to the spherical
geometry.

The forcing parameters for which the spherical drop is unstable are
organized into tongues.  The effect of viscosity is to raise, to
smooth, and to distort the instability tongues, both with increasing
$\nu$ and also with increasing spherical wavenumber $\ell$.
Our computations have demonstrated the appropriate scaling
for the critical oscillation frequency and amplitude with viscosity,
substantially reducing the large parameter space of the problem.

The nonlinear problem is fully three-dimensional and must be treated
numerically. In our companion paper \citep{Ali2}, we will describe and
analyse the patterns corresponding to various values of $\ell$
that we have computed by forcing a viscous drop at appropriate frequencies. 

\subsubsection*{Acknowledgements}
We thank St\'ephan Fauve and David Qu\'er\'e for helpful discussions.

\appendix
\section{Appendix}
\label{appendix}

%%%%%%%%%%%%%%%%%%%%%%%%%%%%%%%%%%%%%%%%%%%%%%%%%%%%%%%%%%%%%%%%%%%%%%%%%%%%
\subsection{Boundary conditions for the two-fluid case}\label{twofl}
%%%%%%%%%%%%%%%%%%%%%%%%%%%%%%%%%%%%%%%%%%%%%%%%%%%%%%%%%%%%%%%%%%%%%%%%%%%%%

We describe here the modifications necessary in order to take into
account the fluid medium surrounding the drop of radius $R$,
occupying either a finite sphere of radius $R_{\rm out}$ or an infinite
domain.  The inner and outer density, dynamic viscosity and
poloidal fields are denoted
by $\rho_j$, $\eta_j$, and $f^{(j)}$ for $j=1,2$.  $\Delta \Psi \equiv
\left[\Psi^{(2)} - \Psi^{(1)}\right]_{r=R}$ denotes the jump of any
quantity $\Psi$ across the interface and applies to all quantities to
its right within a term.  For each spherical harmonic wavenumber
$\ell$ and each Floquet mode $n$, the poloidal fields $f_n^{(j)}$
given in \eqref{flqsolgen} are as follows:
\begin{subequations}
    \begin{align}
f_n^{(1)}(r) & = A_n^{(1)} r^{\ell+1} + B_n^{(1)} r^{\frac{1}{2}}J_{\ell+\frac{1}{2}}(iq_n^{(1)} r) \\
f_n^{(2)}(r) & = A_n^{(2)} r^{\ell+1} + B_n^{(2)} r^{\frac{1}{2}}J_{\ell+\frac{1}{2}}(iq_n^{(2)} r)
+ C_n^{(2)} r^{-\ell} +  D_n^{(2)} r^{\frac{1}{2}}J_{-\left( \ell+\frac{1}{2}\right)}(iq_n^{(2)} r) \end{align}
\label{fnj2fl}
\end{subequations}
where
\begin{equation}
  q_n^{(j)} \equiv \left[\frac{\mu + i(n\omega + \alpha)}{\nu_j}\right]^{1/2}
\end{equation}
The introduction of four more constants requires four additional conditions. 
Two of these are provided by the exterior boundary conditions.
The relations 
\begin{subequations}\begin{align}
  0 &= u_r=\frac{\ell(\ell+1)}{r^2} f \\
  0 &= \nabla_H \cdot\bu_H = \frac{1}{r^2}\frac{\partial}{\partial r} r^2 u_r
  = \frac{1}{r^2}\frac{\partial}{\partial r} \ell(\ell+1) 
  \end{align}\label{eq:uf}\end{subequations}
imply that both $f^{(2)}_n$ and its radial derivative must vanish at $R_{\rm out}$:
\begin{equation}
  0 = f^{(2)}_n(R_{\rm out}) = f^{(2)\prime}_n(R_{\rm out})
\label{eq:ubc2fl}  \end{equation}
If the exterior is infinite, then $A_n^{(2)}=B_n^{(2)}=0$; 
otherwise equations \eqref{eq:ubc2fl} couple the six constants.
Continuity of the velocity at $r=R$, together with \eqref{eq:uf} provides
the remaining two additional conditions:
\begin{equation}
  0 = \Delta f = \Delta f^\prime
%  0 = \left[f_n^{(2)}-f_n^{(1)}\right_{r=R}= \left[f_n^{(2)\prime}-f_n^{(1)\prime}\right]_{r=R}
\label{eq:cu2fl}  \end{equation}
The kinematic condition \eqref{kincondflq} remains unchanged:
\begin{align}
(\mu+i(n\omega+\alpha)) &\zeta_n = \frac{\ell(\ell+1)}{R^2} f_n|_{r=R} 
\label{kincond2fl}\end{align}
since $f$ is continuous across the interface, while the 
tangential stress condition \eqref{bc2flq} becomes
\begin{equation}
0 = \Delta \left[ \eta \left(  \frac{\partial^2}{\partial r^2} - \frac{2}{r}\frac{\partial}{\partial r} + \frac{\ell(\ell+1)}{r^2} \right) f_n \right]
\label{eq:tang2fl}\end{equation}
The pressure jump condition  \eqref{pjcfn} becomes 
\begin{align}
0 = \Delta   \left[ \rho(\mu+i(n\omega + \alpha)) \frac{\partial}{\partial r}\right.
    &\left.- \eta\left(\frac{\partial^3}{\partial r^3} +\frac{2}{r}\frac{\partial^2}{\partial r^2} -\ell(\ell+1)\left(\frac{3}{r^2}\frac{\partial }{\partial r}  -\frac{4}{r^3} \right)\right)f_n \right. &\nonumber\\
&+  \left.\left(\rho g + \sigma \frac{(\ell-1)(\ell+2)}{R^2}\right)\zeta_n 
- \rho \frac{a}{2}(\zeta_{n+1}+\zeta_{n-1})\right]
 \label{pjcfn2fl}
\end{align}
where $\rho$, $\eta$, $\partial^2 f/\partial r^2$ and $\partial^3 f/\partial r^3$ are all discontinuous across the interface.

%%%%%%%%%%%%%%%%%%%%%%%%%%%%%%%%%%%%%%%%%%%%%%%%%%%%%%%%%%%%%%%%%%%%%%%%%%%%%
\subsection{Differentiation relations}
\label{app1}
%%%%%%%%%%%%%%%%%%%%%%%%%%%%%%%%%%%%%%%%%%%%%%%%%%%%%%%%%%%%%%%%%%%%%%%%%%%%%
We express the governing equations in terms of the constants $A_n$, $B_n$, $C_n$, $D_n$ via
\begin{subequations}\begin{align}
   f_n(r)  & =  A_n r^{\ell+1} + B_n r^{\frac{1}{2}}J_{+} + C_n r^{-\ell} + D_n r^{\frac{1}{2}}J_{-} \\   
   \frac{\partial}{\partial r} f_n(r) & =  A_n (l+1)r^\ell + B_n \left( \frac{1}{2} r^{-\frac{1}{2}}J_{+} + iq_n r^{\frac{1}{2}} J_{+}^{\prime} \right) \nonumber\\
   & - C_n \ell r^{-\ell-1} + D_n \left( \frac{1}{2} r^{-\frac{1}{2}}J_{-} + iq_n r^{\frac{1}{2}} J_{-}^{\prime} \right)  \\
   \frac{\partial^2}{\partial r^2} f_n(r) & =  A_n \ell(\ell+1)r^{\ell-1} + B_n \left( -\frac{1}{4} r^{-\frac{3}{2}}J_{+} + iq_n r^{-\frac{1}{2}} J_{+}^{\prime} - q_n^2 r^{\frac{1}{2}} J_{+}^{\prime\prime} \right) \nonumber\\
   & + C_n \ell(\ell+1)r^{-\ell-2} + D_n \left(-\frac{1}{4} r^{-\frac{3}{2}}J_{-} + i q_n r^{-\frac{1}{2}} J_{-}^{\prime} - q_n^2 r^{\frac{1}{2}} J_{-}^{\prime\prime} \right)  \\
   \frac{\partial^3}{\partial r^3} f_n(r) & = A_n (\ell+1)\ell(\ell-1)r^{\ell-2} + B_n \left( \frac{3}{8} r^{-\frac{5}{2}}J_{+} - \frac{3}{4} iq_n r^{-\frac{3}{2}} J_{+}^{\prime}  - \frac{3}{2} q_n^2 r^{-\frac{1}{2}} J_{+}^{\prime\prime} - iq_n^3 r^{\frac{1}{2}} J_{+}^{\prime\prime\prime} \right) \nonumber\\
   & - C_n \ell(\ell+1)(\ell+2)r^{-\ell-3}
   %\nonumber\\ &
   + D_n \left( \frac{3}{8} r^{-\frac{5}{2}}J_{-} - \frac{3}{4} iq_n r^{-\frac{3}{2}} J_{-}^{\prime}  - \frac{3}{2} q_n^2 r^{-\frac{1}{2}} J_{-}^{\prime\prime} - iq_n^3 r^{\frac{1}{2}} J_{-}^{\prime\prime\prime} \right)
\end{align}
\label{derfn}
\end{subequations}
\noindent where $J_+$ and $J_-$ denote $J_{\ell+\frac{1}{2}}$ and $J_{-\left(\ell+\frac{1}{2}\right)}$, respectively, to be evaluated at $iq_n^{(j)}R$.
To evaluate the derivatives of the Bessel functions, we
use the recurrence relations:
\begin{subequations}
    \begin{align}
    J^\prime_\nu(z)& = \frac{1}{2} \left( J_{\nu-1}(z) -  J_{\nu+1}(z) \right) \\
    J^{\prime\prime}_\nu(z)& = \frac{1}{4} \left( J_{\nu-2}(z) - 2J_\nu(z) + J_{\nu+2}(z) \right) \\
    J^{\prime\prime\prime}_\nu(z)& = \frac{1}{8} \left( J_{\nu-3}(z) - 3J_{\nu-1}(z) + 3J_{\nu+1}(z) - J_{\nu+3}(z) \right) 
    \end{align}
\label{derbf}
\end{subequations}
For the two-fluid case, 
we express the seven conditions \eqref{eq:ubc2fl}, \eqref{eq:cu2fl}, \eqref{kincond2fl}, \eqref{eq:tang2fl} and \eqref{pjcfn2fl} in terms of the constants
$A^{(j)}_n$, $B^{(j)}_n$, $C^{(j)}_n$, $D^{(j)}_n$, and $\zeta_n$. 
For the single-fluid case, we express the 
three conditions \eqref{kincondflq}, \eqref{bc2flq}, \eqref{pjcfn}
in terms of $A_n$, $B_n$, and $\zeta_n$.
Omitting the pressure jump condition leads to a
$6\times 6$ (finite outer sphere), $4\times 4$ (infinite outer sphere),
or $2\times 2$ (single-fluid) system which can be inverted
to obtain values for all of the constants as multiples of $\zeta_n$.
The pressure jump condition is then a Floquet problem in $\{\zeta_n\}$,
solved as described in section \ref{Floquet Solution}.

% Pour finir l'interligne de 1,5
%\end{onehalfspace}

%----------------------------------------
% Pour la bibliographie
%----------------------------------------
%\clearpage

%\bibliographystyle{jfm}
%\bibliography{sphere}

\begin{thebibliography}{25}
\expandafter\ifx\csname natexlab\endcsname\relax\def\natexlab#1{#1}\fi
\def\au#1{#1} \def\ed#1{#1} \def\yr#1{#1}\def\at#1{#1}\def\jt#1{\textit{#1}}
  \def\bt#1{#1}\def\bvol#1{\textbf{#1}} \def\vol#1{#1} \def\pg#1{#1}
  \def\publ#1{#1}\def\arxiv#1{#1}\def\org#1{#1}\def\st#1{\textit{#1}}

\bibitem[Abramowitz \& Stegun(1965)]{AS65}
{\sc \au{Abramowitz, M.} \& \au{Stegun, I.A.}} \yr{1965} {\em Handbook of
  Mathematical Functions\/}.  \publ{Dover Publications}.

\bibitem[Batson {\em et~al.\/}(2013)Batson, Zoueshtiagh \& Narayanan]{Batson}
{\sc \au{Batson, W.}, \au{Zoueshtiagh, F.} \& \au{Narayanan, R.}} \yr{2013}
  \at{The \protect{Faraday} threshold in small cylinders and the sidewall
  non-ideality}.  \jt{J. Fluid Mech.}  \bvol{729},  \pg{496--523}.

\bibitem[Bechhoefer {\em et~al.\/}(1995)Bechhoefer, Ego, Manneville \&
  Johnson]{BEMJ1995}
{\sc \au{Bechhoefer, J.}, \au{Ego, V.}, \au{Manneville, S.} \& \au{Johnson,
  B.}} \yr{1995}  \at{An experimental study of the onset of parametrically
  pumped surface waves in viscous fluids}.  \jt{J. Fluid Mech.}  \bvol{288},
  \pg{325--350}.

\bibitem[Benjamin \& Ursell(1954)]{BU1954}
{\sc \au{Benjamin, T.~B.} \& \au{Ursell, F.}} \yr{1954}  \at{The stability of
  the plane free surface of a liquid in vertical periodic motion}.  \jt{Proc.
  R. Soc. Lond.}  \bvol{225},  \pg{505--515}.

\bibitem[Brunet \& Snoeijer(2011)]{BS2011}
{\sc \au{Brunet, P.} \& \au{Snoeijer, J.~H.}} \yr{2011}  \at{Star-drops formed
  by periodic excitation and on an air cushion--a short review}.  \jt{Eur.
  Phys. J. Special Topics}  \bvol{192}~(1),  \pg{207--226}.

\bibitem[Chandrasekhar(1961)]{Chan1961}
{\sc \au{Chandrasekhar, S.}} \yr{1961} {\em Hydrodynamic and Hydromagnetic
  Stability\/}.  \publ{Oxford University Press}.

\bibitem[Cummings \& Blackburn(1991)]{CB1991}
{\sc \au{Cummings, D.~L.} \& \au{Blackburn, D.~A.}} \yr{1991}  \at{Oscillations
  of magnetically levitated aspherical droplets}.  \jt{J. Fluid Mech.}
  \bvol{224},  \pg{395--416}.

\bibitem[Ebo-Adou {\em et~al.\/}(2019)Ebo-Adou, Tuckerman, Shin, Chergui \&
  Juric]{Ali2}
{\sc \au{Ebo-Adou, A.}, \au{Tuckerman, L.S.}, \au{Shin, S.}, \au{Chergui, J.}
  \& \au{Juric, D.}} \yr{2019}  \at{Faraday instability on a sphere: Numerical
  simulation}.  \jt{J. Fluid Mech.}  \bvol{870},  \pg{433--459}.

\bibitem[Falc\'on {\em et~al.\/}(2009)Falc\'on, Falcon, Bortolozzo \&
  Fauve]{Fal2009}
{\sc \au{Falc\'on, C.}, \au{Falcon, E.}, \au{Bortolozzo, U.} \& \au{Fauve, S.}}
  \yr{2009}  \at{Capillary wave turbulence on a spherical fluid surface in low
  gravity}.  \jt{Europhys. Lett.}  \bvol{86},  \pg{14002}.

\bibitem[Futterer {\em et~al.\/}(2013)Futterer, Krebs, Plesa, Zaussinger,
  Hollerbach, Breuer \& Egbers]{futterer2013sheet}
{\sc \au{Futterer, B.}, \au{Krebs, A.}, \au{Plesa, A.~C.}, \au{Zaussinger, F.},
  \au{Hollerbach, R.}, \au{Breuer, D.} \& \au{Egbers, C.}} \yr{2013}
  \at{Sheet-like and plume-like thermal flow in a spherical convection
  experiment performed under microgravity}.  \jt{J. Fluid Mech.}  \bvol{735},
  \pg{647--683}.

\bibitem[Kelvin(1863)]{Kel1863}
{\sc \au{Kelvin, Sir~Lord}} \yr{1863}  \at{Elastic spheroidal shells and
  spheroids of incompressible liquid}.  \jt{Philos. Trans. R. Soc. London}
  \bvol{153},  \pg{583}.

\bibitem[Kumar(1996)]{K1996}
{\sc \au{Kumar, K.}} \yr{1996}  \at{Linear theory of {F}araday instability in
  viscous fluids}.  \jt{Proc. R. Soc. Lond.}  \bvol{452},  \pg{1113--1126}.

\bibitem[Kumar \& Tuckerman(1994)]{KT1994}
{\sc \au{Kumar, K.} \& \au{Tuckerman, L.~S.}} \yr{1994}  \at{Parametric
  instability of the interface between two fluids}.  \jt{J. Fluid Mech.}
  \bvol{279},  \pg{49--68}.

\bibitem[Lamb(1932)]{Lamb1932}
{\sc \au{Lamb, H.}} \yr{1932} {\em Hydrodynamics\/}.  \publ{Cambridge
  University Press}.

\bibitem[Marston(1980)]{Mars1980}
{\sc \au{Marston, P.~L.}} \yr{1980}  \at{Shape oscillation and static
  deformation of drops and bubbles driven by modulated radiation
  stresses-—theory}.  \jt{J. Acoust. Soc. Am.}  \bvol{67}~(1),  \pg{15--26}.

\bibitem[Miller \& Scriven(1968)]{MS1968}
{\sc \au{Miller, C.~A.} \& \au{Scriven, L.~E.}} \yr{1968}  \at{{T}he
  oscillations of a fluid droplet immersed in another fluid}.  \jt{J. Fluid
  Mech.}  \bvol{32},  \pg{417--435}.

\bibitem[Prosperetti(1980)]{Pros1980}
{\sc \au{Prosperetti, A.}} \yr{1980}  \at{{N}ormal-mode analysis for the
  oscillations of a viscous liquid drop in an immiscible liquid}.  \jt{J. Mec.}
   \bvol{19},  \pg{149--182}.

\bibitem[Rayleigh(1879)]{Rayl1879}
{\sc \au{Rayleigh, Lord}} \yr{1879}  \at{{O}n the capillary phenomena of jets}.
   \jt{Proc. R. Soc. Lond.}  \bvol{29},  \pg{71--97}.

\bibitem[Reid(1960)]{Reid1960}
{\sc \au{Reid, W.~H.}} \yr{1960}  \at{{T}he oscillations of a viscous liquid
  drop}.  \jt{Q. Appl. Math.}  \bvol{18},  \pg{86--89}.

\bibitem[Shen {\em et~al.\/}(2010)Shen, Xie \& Wei]{SXW2010}
{\sc \au{Shen, C.~L.}, \au{Xie, W.~J.} \& \au{Wei, B}} \yr{2010}
  \at{Parametrically excited sectorial oscillation of liquid drops floating in
  ultrasound}.  \jt{Phys. Rev. E}  \bvol{81}~(4),  \pg{046305}.

\bibitem[Trinh \& Wang(1982)]{TW1982}
{\sc \au{Trinh, E.} \& \au{Wang, T.~G.}} \yr{1982}  \at{Large-amplitude free
  and driven drop-shape oscillations: experimental observations}.  \jt{J. Fluid
  Mech.}  \bvol{122},  \pg{315--338}.

\bibitem[Trinh {\em et~al.\/}(1982)Trinh, Zwern \& Wang]{TZW1982}
{\sc \au{Trinh, E}, \au{Zwern, A} \& \au{Wang, T.~G.}} \yr{1982}  \at{An
  experimental study of small-amplitude drop oscillations in immiscible liquid
  systems}.  \jt{J. Fluid Mech.}  \bvol{115},  \pg{453--474}.

\bibitem[Tsamopoulos \& Brown(1983)]{TB1983}
{\sc \au{Tsamopoulos, J.~A.} \& \au{Brown, R.~A.}} \yr{1983}  \at{{N}onlinear
  oscillations of inviscid drops and bubbles}.  \jt{J. Fluid Mech.}
  \bvol{127},  \pg{514--537}.

\bibitem[Vega {\em et~al.\/}(2001)Vega, Knobloch \& Martel]{VKM2001}
{\sc \au{Vega, J.~M.}, \au{Knobloch, E.} \& \au{Martel, C.}} \yr{2001}
  \at{Nearly inviscid {F}araday waves in annular containers of moderately large
  aspect ratio}.  \jt{Physica D}  \bvol{154},  \pg{313--336}.

\bibitem[Wang {\em et~al.\/}(1996)Wang, Anilkumar \& Lee]{WAL1996}
{\sc \au{Wang, T.~G.}, \au{Anilkumar, A.~V.} \& \au{Lee, C.~P.}} \yr{1996}
  \at{Oscillations of liquid drops: results from \protect{USML-1} experiments
  in space}.  \jt{J. Fluid Mech.}  \bvol{308},  \pg{1--14}.

\end{thebibliography}

\end{document}